\documentstyle[10pt,aaspp4]{article}
\newcommand\mathnew{\mathsurround=0pt}
\def\simov#1#2{\lower .5pt\vbox{\baselineskip0pt \lineskip-.5pt
        \ialign{$\mathnew#1\hfil##\hfil$\crcr#2\crcr\sim\crcr}}}
\newcommand\simg{\mathrel{\mathpalette\simov >}}
\newcommand\siml{\mathrel{\mathpalette\simov <}}
\date{today}
\begin{document}
\begin{center}
\title{IMPACT OF RELATIVISTIC FIREBALLS ON EXTERNAL MATTER:\\ 
NUMERICAL MODELS OF COSMOLOGICAL GAMMA-RAY BURSTS}
\author{A. Panaitescu, L. Wen\altaffilmark{1}, P. Laguna\altaffilmark{2} \& 
P. M\'esz\'aros\altaffilmark{2}}
\affil{Department of Astronomy \& Astrophysics, \\
Pennsylvania State University, University Park, PA 16802}
\altaffiltext{1}{now Physics Department, M.I.T., Blg. 6-110, Cambridge, MA 02139}
\altaffiltext{2}{also Center for Gravitational Physics and Geometry,
Pennsylvania State University}
\bigskip
\end{center}

\begin{abstract}

We numerically model the interaction between an expanding fireball and a
stationary external medium whose density is either homogeneous or varies
with distance as a power-law. The evolution is followed until most of the 
fireball kinetic energy is converted into internal energy. The density, 
pressure and flow Lorentz factor profiles are shown at different stages, 
including shock and rarefaction wave reflections, for a fireball 
of initial bulk Lorentz factor $\Gamma=100$, both in the adiabatic and 
non-adiabatic (radiative) regimes. For cooling times shorter than the 
dynamic time, bolometric light-curves are computed for values of $\Gamma
=50,~100$ and $200$. We compare the numerical light-curves with analytic 
results, and find that for a homogeneous external medium there is a simple 
scaling relationship among light-curves obtained for different parameters. 
The light-curves for power-law external densities are similar in shape to
those in the homogeneous case. We discuss the implications of a comparison
of the results with observed Gamma-Ray Burst time histories. 

\end{abstract}

\keywords{gamma-rays: bursts - hydrodynamics - methods: numerical - relativity - 
          shock waves}

\section{Introduction}
\label{introduction}

Although the cosmological origin of Gamma-Ray Bursts (GRB) is still contested, 
there is good evidence to support it. The limits on the isotropy of their 
spatial distribution from the Burst and Transient Source Experiment (BATSE) 
put increasingly strong constraints on alternative hypotheses 
(\cite{fm95}; \cite{mee96}), the lack of evidence for repetition is becoming significant
(\cite{teg96}), there is possible evidence for correlation with Abell rich 
clusters (\cite{kopi96}) and cosmological time-dilation (\cite{nea95}; \cite{nea96}), 
as well as spectral softening effects may have been found (\cite{mal95},
see however \cite{mit96}). In this paper we proceed on the assumption 
that GRBs are cosmological in origin, and therefore result in a relativistically
expanding fireball that produces a burst when shocks develop in the optically
thin phase of the expansion (see \cite{me95} for a review). Internal shocks 
occur at smaller radii, are marginally relativistic in the comoving frame,
and can give rise to arbitrarily complicated light-curves whose duration is
determined by the details of the energy release mechanisms (\cite{rm94};
\cite{pm96}). On the other hand, external shocks arising at larger radii from
interaction with an external medium (EM) are extremely relativistic in the
forward blast wave that moves into EM (\cite{mrp94}; \cite{sapi95}).
In this case, the burst duration is determined, for reasonable efficiencies,
by the energetics of the source ($E$, $\eta$) and the density of the
external medium ($n$), via the Doppler time-delays of the light arriving
from different parts of the light cone (\cite{re66}; \cite{mr93}). 

Numerical simulations of relativistic jets with Lorentz factor $\sim 22$ have 
been published by \cite{marti95} and shock propagation over a limited region,
generated from flows with Lorentz factors up to $\sim 2000$ have been used as tests 
(\cite{marti96}; \cite{romero96}), but numerical hydrodynamic simulations
of shocks with very high ($\Gamma \simg 50$) Lorentz factors and their 
propagation over times that are several orders of magnitude longer than the 
system's crossing time, do not appear to have been published so far.
This problem is of interest not only for GRB 
but also for relativistic pulsar winds interacting with a surrounding 
nebula, and for AGN jets, in which the intraday variability reported at radio 
wavelengths by \cite{hw96}, and at TeV energies by \cite{gai96},
may require bulk Lorentz factors in excess of 50--100. For GRB, such numerical
calculations of the time variation of the dynamical variables
are most urgently needed for computing
accurate light-curves from external shocks.
A discussion of BATSE 3B pulse profiles has been presented by Fenimore, Madras
\& Nayakshin (1996), who model the burst as an energy release at constant power 
from a relativistically expanding surface. 
In this paper, we calculate numerically the 
energy release as a continuum of events from a relativistic fluid 
subject to shocks, assuming a radiating power that is proportional to the amount of 
internal energy stored in each elementary volume element.
We discuss the dynamics of the forward and
reverse shocks, including the phenomena of shock and rarefaction wave 
reflection, over the range of radii within which the ejecta randomize and 
radiate most of its initial kinetic energy. We compute numerical light-curves 
for initial bulk Lorentz factors $\Gamma =50$ to 200, in which the 
fireball ejecta interacts with a homogeneous or a power-law external medium,
we compare these numerical light-curves with observations,
and discuss the implications for GRB models.

\section{Initial Conditions in the Expanding Fireball}
\label{free expansion}

The initial conditions of an impulsive fireball are: the total
initial energy released $E$, the dimensionless entropy or initial random 
Lorentz factor $\eta=E/Mc^2$ (or, equivalently, the amount of baryonic mass entrained in 
the fireball $M = E/\eta c^2 = 5.6\times10^{-6}E_{51}\eta_2^{-1}\;M_{\odot}$)
and the size $r_0$ of the region where the energy $E$ is released.
These parameters determine entirely the initial free expansion stage of
the fireball (see M\'esz\'aros, Laguna \& Rees 1993).
Conversion of the energy $E$ into radiation
leads to a typical fluence at Earth of $F = 10^{-6}E_{51}D_{28}^{-2} 
\; {\rm ergs/cm}^2$, at a distance $D=10^{28}D_{28}$ cm. In the beginning of the
expansion, the fireball is optically thick with respect to pair formation and/or Compton 
scattering, out to radii as large as $r\sim 10^{13}$ cm in the {\it source 
frame} attached to the fixed origin point of the explosion (two other frames
in use are the {\it comoving frame} attached to moving portions of the flow, 
and the {\it detector} or observer {\it frame}). As it expands, the fireball converts 
its internal energy into bulk kinetic energy. 
By the time it becomes optically thin, most of the initially available 
energy has been turned into kinetic energy, which can be reconverted 
into random energy and radiation via shocks. The above parameters and the
characteristic particle density $n$ of the external medium determine the {\sl
deceleration radius} around which this occurs, given by the condition $E \sim  
(4\pi/3) r_{dec}^3 n \eta^2 m_p c^2$ (see eq. [\ref{deceleration}]).
For computational economy, we do not start modeling the fireball--EM interaction
from the very beginning of the fireball's expansion, but only after it has 
traveled for a few tenths of the deceleration radius.
Before this stage is reached, the mass of the swept up EM is a
small fraction of the mass within the deceleration radius; therefore any back-reaction 
caused by the interaction with the EM may be neglected 
and the fireball can be considered in free expansion.

During the free expansion phase most of the fireball's mass becomes concentrated in a 
thin shell near the outer edge of the expanding gas (M\'esz\'aros et al. 1993), 
and we will consider that 
practically all baryons are in this homogeneous shell, delimited by two discontinuities.
The shell's Lorentz factor 
initially grows as $r$, and typically reaches a saturation value $\Gamma\sim 
\eta$ at the {\sl saturation radius} $r_s= \eta r_0= 10^8\;\eta_2\;r_{0,6}\;{\rm 
cm}$, after which the cooled fireball coasts at constant Lorentz factor. 
The lab frame ejecta shell width is constant $\Delta\simeq r_0$ until the {\sl 
broadening radius} $r_b=\eta^2\,r_0=10^{10}\,\eta_2^2\,r_{0,6}$ cm is reached, 
after which the shell's width increases linearly with radius: $\Delta \sim
r/\eta^2$. The optical depth for Thomson scattering $\tau (r)= E\,\kappa\,/
4\,\pi\,r^2\,c^2\,\eta$ has the value $\tau (r_s)= 3.5\times10^{10}\;E_{51}
\,r_{0,6}^{-2}\,\eta_2^{-3}$ at $r_s$ ($\kappa=0.40\;{\rm cm^2/g}$ is the mass 
absorption coefficient for Thomson scattering) and is larger than 1 as long as 
$\log \eta_2 + 0.67\,\log r_{0,6} - 0.33\,\log E_{51} < 3.5$ .
The durations of the observed GRBs require that $\eta \siml 10^3$,
so that for $r_0 \siml 5\times10^9\;{\rm cm}$ and $E=10^{51}$ ergs,
the Lorentz factor reaches saturation before optical thinness.
Therefore the {\sl thinning radius} is $r_t=(E\,\kappa\,/4\,\pi\,c^2\,
\eta)^{1/2}= 1.9\times10^{13}\,E_{51}^{1/2}\,\eta_2^{-1/2}\;{\rm cm}$.
If $r_t$ is reached before $r_s$,
the shell's Lorentz factor at $r_t$ is less than $\eta$, but 
subsequent Compton drag accelerate the shell up to a Lorentz factor
comparable with $\eta$ (see M\'esz\'aros et al. 1993).
From here on we will denote by $\Gamma\;(\sim \eta)$ the maximum Lorentz 
factor the shell reaches at the end of the acceleration phase.
Since the deceleration radius (see eq. [\ref{deceleration}]) is much
larger than both $r_s$ and $r_t$, the shell travels at constant Lorentz
factor $\Gamma$ before the interaction with the EM cannot be neglected any 
longer and must be taken into account. For completeness, we note that $r_t$ is 
reached before $r_b$ if $\log \eta_2+0.4\,\log r_{0,6}-0.2\,\log E_{51} > 1.3$ ,
and that the broadening radius is always larger than the saturation radius.

The adiabatic index of the mixture protons + electrons + photons is determined 
by the component that gives the largest contribution to the total pressure. The 
ratio of the photon--to--electron (or proton) partial pressures is 
$P_{photon}/P_{electron}=aVT^3/3Nk_B$, where N is the total number of electrons 
(baryons) and  V is the comoving volume of the ejecta shell. Since the
adiabatic index is initially $\hat\gamma=4/3$ (the ejecta are radiation dominated, both in 
density and pressure), the product $VT^3$ is constant and the pressure ratio 
does not change. This means that the ejecta remain radiation dominated and 
the adiabatic index is 4/3 as long as the electrons and photons are coupled, 
i.e. until the thinning radius $r_t$ is reached. After $r_t$, the mixture 
electrons + protons is already cold (since $r_t > r_s$) and the expansion 
is characterized by the adiabatic index 5/3. Taking into account the change in 
the adiabatic index after decoupling, the ejecta comoving density and pressure 
are given by:
\begin{equation}
\rho = \left\{ \begin{array}{ll}
\rho_0(r_0/r)^3  & r < r_s \\
\rho_0\eta^{-1}(r_0/r)^2  &  r_s < r < r_b \\
\rho_0\eta(r_0/r)^3  & r_b < r
\end{array} \right.
\label{density} 
\end{equation}
\begin{equation}
p = \left\{ \begin{array}{ll}
p_0(r_0/r)^4  & r < r_s \\
p_0\eta^{-4/3}(r_0/r)^{8/3} &  r_s < r < r_t \\
p_0\eta^{-4/3}(r_t r_0^4/r^5)^{2/3} & r_t < r < r_b \\
p_0\eta^2 (r_t^2 r_0^{13}/r^{15})^{1/3} & r_b < r
\end{array} \right.
\qquad \quad
p = \left\{ \begin{array}{ll}
p_0(r_0/r)^4  & r < r_s \\
p_0\eta^{-4/3}(r_0/r)^{8/3} & r_s < r < r_b \\
p_0\eta^{4/3}(r_0/r)^4 & r_b < r < r_t \\
p_0\eta^{4/3}(r_t r_0^4/r^5) & r_t < r
\end{array} \right. ,
\label{pressure} 
\end{equation}
\qquad \qquad \qquad if $r_t < r_b$ \qquad \qquad \qquad \qquad and \qquad 
\qquad \qquad if $r_t > r_b$ \\
where  $\rho_0= 8.8 \times 10^8 \,E_{51}\,r_{0,6}^{-3}\,\eta_2^{-1}\,{\rm g/cm^3}$
and $p_0= 2.1 \times 10^{28}\,E_{51}^{5/4}\,r_{0,6}^{-15/4}\,\eta_2^{-1}\,
{\rm dyne/cm^2}$ are the baryonic density and pressure at $r=r_0$.

When the ejecta shell interacts with the EM, a forward shock (blast 
wave) sweeps up the EM, heating and collecting 
it in a dense sub-shell between the shock and the contact discontinuity (CD). 
A reverse shock propagates backwards into the fireball ejecta shell, 
compressing and heating the ejecta between the reverse shock and the CD.
These shocks convert an increasing fraction of the ejecta bulk kinetic energy 
into internal energy. The deceleration of the fireball is important when the 
energy stored in the shocked material is a substantial fraction of the 
fireball's initial kinetic energy ($\sim E$). The random (or thermal) Lorentz 
factor of the shocked EM protons is comparable to the bulk Lorentz factor of 
the upstream ejecta (see eq. [3] and [10] in Blandford \& McKee 1976). 
Therefore, for a homogeneous EM, the deceleration radius is given by
\begin{equation}
r_{dec} = (3E/4\pi \eta^2 n m_p c^2)^{1/3} 
\sim 2.5\times 10^{16} \, (E_{51}/n_0)^{1/3} \eta_2^{-2/3}~{\rm cm }~,
\label{deceleration}
\end{equation}
where $n_0$ is the EM number density in ${\rm cm}^{-3}$. Since the shell is
ultra-relativistic, the source-frame deceleration time-scale (or hydrodynamic
time-scale) is given by $t_{dec}=r_{dec}/c$ . Note that the deceleration 
radius is orders of magnitude larger than the saturation and thinning radii, 
so that at $r_{dec}$ the ejecta are very rarefied and optically thin. 
In the absence of the deceleration caused by the EM, 
the nominal lab frame shell width, comoving density and 
pressure at the deceleration radius, derived from equations 
(\ref{density}), (\ref{pressure}) and (\ref{deceleration}), would be: 
$\Delta_{dec} = 84\;E_{51}^{1/3}\,\eta_2^{-8/3}\,n_0^{-1/3}\,c\; {\rm s}$, 
$\rho_{dec}^{shell} = 5.0\;\eta_2^2\,n_0\,c^{-2}\; {\rm ergs/cm^3}$, and 
$p_{dec}^{shell} = 1.5\times10^{-15}\;E_{51}^{-1/12}\,r_{0,6}^{7/12}\;\eta_2^4
\;n_0^{5/3}\; {\rm dynes/cm^2}$ if $r_t < r_b$ or $p_{dec}^{shell} = 
1.9\times10^{-14}\;E_{51}^{1/12}\,r_{0,6}^{1/4}\,\eta_2^{19/6}\, n_0^{5/3}\;
{\rm dynes/cm^2}$ if $r_t > r_b$. Throughout this paper, we will use the 
comoving rest mass energy density $\rho\,c^2$, rather than the comoving mass
density $\rho$, which has the advantage of showing whether the material is cold 
($p < \rho c^2$) or hot ($p > \rho c^2$) . 
The width and the density and pressure inside the shell at the point 
from which we start modeling the interaction with the EM can be found from 
their nominal values at the deceleration radius, using the scaling equations 
$p \propto r^{-5}$, $\rho \propto r^{-3}$ and $\Delta \propto r$ (from 
eqs. [\ref{density}] and [\ref{pressure}]).
 
As viewed by the observer, the kinematics of the shell is similar to that of the
relativistically expanding radio source described by Rees (1966). Due to the
strong beaming effects, the observer receives radiation mainly from a spherical 
cap of opening angle $\approx \Gamma^{-1}$. This cap has an apparent speed 
towards the observer of $\simeq 2\,\Gamma^2\,c$, so the effects of the 
deceleration of the ejecta shell are observed on a time-scale
\begin{equation}
T_{burst} \equiv t_{dec}\,/\,2\,\Gamma^2 = 42 \;\,E_{51}^{1/3}\;\Gamma_2^{-8/3}\;
n_0^{-1/3}\; {\rm s}.
\label{duration}
\end{equation}
Here we use lower case $t$ to denote the lab frame (or source) time 
and upper case $T$ for the detector time.
Since the Lorentz factor of the shocked material will be less than the initial 
Lorentz factor $\Gamma$ of the colliding shell, due to deceleration, the actual 
duration of the burst will be somewhat longer than predicted by equation 
(\ref{duration}). This means  that $\Gamma$ should be in the range $10^2-10^3$ in 
order to explain the observed burst durations.

\section{Numerical Results and Light Curves from Shocks with $\Gamma=100$ in a
         Homogeneous External Medium}

Analytical approximations for the dynamics of highly relativistic shocks in 
the adiabatic limit and neglecting gradients of the physical quantities, were
presented by \cite{blamc76} and for GRB by M\'esz\'aros \& Rees (1993), 
\cite{mlr93}, and by Sari \& Piran (1995), \cite{sanapi96}. 
The latter consider details of the dynamics for 
situations where the shocked material is hot and the reverse 
shock is either Newtonian or ultra-relativistic, as seen from the colliding 
shell's frame. Numerical results on relativistic shocks 
(mainly related to AGN jets or supernova remnants) have so far been restricted 
to cases with bulk Lorentz factors $\Gamma \siml 5-10$. Our goal is to 
model numerically the deceleration of the shell and the dynamics of the 
reverse shock for any value of its Lorentz factor,
including the trans-relativistic case of considerable importance
for bursts arising from internal shocks or from reverse shocks (in the external
shock model).

For our numerical simulations we have developed a hybrid code based on 
standard Eulerian finite difference techniques in most of the computational 
domain, and a Glimm algorithm including an exact Riemann Solver in regions where discontinuities 
are present. The code was tested using standard shock tube problems and 
planar and spherical relativistic shock reflection problems. An analysis of 
the errors and convergence rates shows that the quality of our results is comparable
to that in the Piecewise Parabolic Method (\cite{marti96}; \cite{romero96}). 
A description of this code and details of the
tests performed to calibrate it are given in \cite{wen96}. The average mass
and total energy relative errors during the runs done to obtain the numerical results 
described below, were less than 3\%.

In this section we outline the calculations for an ejecta shell colliding with a
homogeneous EM, for a representative set of parameters $\Gamma=100$, $E=10^{51}$
ergs, $r_0=10^8$ cm, $n=1\; {\rm cm^{-3}}$. Results for different parameters 
and the effect of an inhomogeneous EM are presented in \S 4.  
The evolution of the interaction shell--EM is followed from $t=0.4\;t_{dec}$, at which 
point the physical parameters characterizing the shell are $\Delta=33.5\;c\;
{\rm s}=1.01\times 10^{12}\;{\rm cm}$, $p=5.74\times10^{-12}\;{\rm dyne/cm}^2$
and $\rho=78.3\;c^{-2}\;{\rm ergs/cm}^3= 8.70\times 10^{-20}\;{\rm g/cm}^3$.
Deceleration effects should be negligible up to this point since only 6.4\,\% of 
the EM mass within the deceleration radius has been swept up. We consider two
simple cases: the adiabatic case (no energy leaves the system) and a simplified
radiative collision case where it is assumed that a fraction 1/250 
of the internal energy is radiated by the system every
$\delta t=10^{-3} t_{dec}$. We refer to this recipe for energy release 
as the ``$0.25\;t_{dec}$ energy release time-scale".

Figure 1 compares the density, pressure and Lorentz factor $\gamma$ (we use
the lower case letter for the flow Lorentz factor after the free expansion
phase) inside the 
composite structure shocked shell--shocked EM, in the adiabatic
and radiative cases. The abscissae give the (Eulerian) position relative to
the CD that separates the ejecta and the shocked EM shells. Therefore, 
in all graphs, the CDs at different times between $t=0.5\;t_{dec}$ 
and $t=0.9\;t_{dec}$, in steps of $0.1\;t_{dec}$, are coincident. 
Each density profile has the same structure, showing from left to right: the 
unshocked fireball, the reverse shock propagating radially inward into the
fireball, the shocked fireball material (condensed by a factor $\siml 10$), 
the CD between the inner (fireball) and outer (EM) shells, 
the shocked EM ($\approx 100$ times denser than the unshocked EM) and the blast 
wave (or forward shock) that propagates radially outward in the EM. The 
post-shock pressure and density satisfy the strong shock equations in 
\cite{blamc76}. One sees that $\rho c^2 > p$ in the inner shell (therefore this shell
is cold) and that $p > \rho c^2$ in the outer shell (the shocked EM is hot). If no energy 
leaves the structure, the internal energy of the shocked regions increases as 
more and more heated material accumulates between the two shocks, and accounts 
for the lost kinetic energy. Note that, in the adiabatic interaction, the 
shocked EM density is almost constant throughout the shell and slowly decreases 
in time, while in the non-adiabatic case the outer shell density is larger 
before the CD than behind the forward shock, and increases 
in time.  In the adiabatic case, the thermal Lorentz factor of the shocked EM 
changes little with position and decreases from $\approx 32$ at 
$t=0.5\;t_{dec}$ to $\approx 26$ at $t=0.9\;t_{dec}$. In the non-adiabatic 
case, the same thermal Lorentz factor is lower before the CD
($\approx 26$ at $t=0.5\;t_{dec}$, $\approx 10$ at $t=0.9\;t_{dec}$) than 
behind the forward shock ($\approx 30$ at $t=0.5\;t_{dec}$, $\approx 20$ at 
$t=0.9\;t_{dec}$), since material shocked earlier had more time to radiate 
its internal energy. By the time the structure reaches $0.92\;r_{dec}$, the 
reverse shock has swept up all the ejecta shell gas (in both cases).
The shock crossing time $t_{cross}^{shock}=0.52\;t_{dec}=4.4\times 10^5$ s is
within a factor 2 to the crossing time calculated by Sari \& Piran (1995). 
The reverse shock velocity $\beta_{sh}^{co}$ in the frame of the unshocked fluid
(which moves in the lab frame at $\beta=\sqrt{\Gamma^2-1}/\Gamma$) is related to the
lab frame reverse shock velocity $\beta_{sh}^{lab}$ by
$\beta-\beta_{sh}^{lab}\simeq \beta_{sh}^{co}\,\Gamma^{-2}/(1-\beta_{sh}^{co})$,
from which  $\beta_{sh}^{co}=\Gamma^2 (\beta-\beta_{sh}^{lab})/
[1+\Gamma^2 (\beta-\beta_{sh}^{lab})]$ (the speed of light is set to 1). 
In the lab frame, $\beta-\beta_{sh}^{lab}
=\Delta/t_{cross}^{shock} \simeq 7.7\times 10^{-5}$, therefore $\beta_{sh}^{co}
=0.43$ (Lorentz factor $\gamma_{sh}^{co} \simeq 1.10$). The reverse 
shock is trans-relativistic (or quasi-Newtonian) in the ejecta shell rest frame.

After the reverse shock has crossed the fireball ejecta, a rarefaction wave 
propagates forward into the shocked inner shell.
This wave travels in the comoving frame 
with the local sound speed, which in the adiabatic case is
$\beta_w^{co}=c_s=\sqrt{\hat\gamma p/h} 
\simeq 0.21$, where $h$ is the comoving enthalpy density. 
In the lab frame, the wave's velocity relative to the 
CD (which moves at $\gamma_{cd} \simeq 66$) is $\beta_w^{lab}-
\beta_{cd} \simeq \beta_w^{co} \gamma_{cd}^{-2}/(1+\beta_w^{co}) \simeq 
4.0\times 10^{-5}$.  When the reverse shock crosses the inner shell, the width
of this shell is $\Delta_{in} \simeq 6.4$ light-seconds, therefore the wave 
lab crossing time is $t_{cross}^{wave}=\Delta_{in}/(\beta_w^{lab}-\beta_{cd})
= 1.6\times10^5\,{\rm s} \simeq 0.19\;t_{dec}$. In the non-adiabatic case,
the wave's speed relative to the CD is $\beta_w^{lab}-\beta_{cd} \simeq
3.7\times10^{-5}$ and the inner shell's thickness is $\Delta_{in} \simeq 3.4$
light-seconds, therefore the wave lab crossing time is  $t_{cross}^{wave}=
9.0\times10^4\,{\rm s} \simeq 0.11\;t_{dec}$. Figure 2 shows 
that after the rarefaction wave has crossed the inner shell,  
a second reverse shock forms and propagates in the now 
rarefied material behind the CD (this is easier to see in the 
Lorentz factor graphs, which show the shocked material being decelerated across
the shock). 
Note that when energy is released from the system the shocked EM is denser.
In the adiabatic interaction, the shocked EM behind the blast wave 
moves faster than that next to the CD, 
and a rarefaction wave develops in the outer shell. 
This is due to the fact that in the adiabatic interaction the shocked EM is hotter
and a fraction of the internal energy is re-converted into kinetic, accelerating the 
forward shock more efficiently than in the non-adiabatic case. This effect can be seen 
in all graphs in figures 2 and 3, by comparing the forward shock position at the
same times.
In both situations (adiabatic or non-adiabatic), the thermal Lorentz factor in 
the outer shell decreases in time, due to the 
expansion of this shell (and the energy release, in the non-adiabatic case).
Figure 3 shows the evolution of the 
structure until $t=2.0\;t_{dec}$. At $t=1.92\;t_{dec}$ all the fireball's gas
has been compressed behind the second reverse shock, and a new rarefaction wave 
propagates forward in the shell. At $t=2.0\;t_{dec}$, when we stop the simulation, 
the whole structure extends over $\approx 1000$ light-seconds and consists of 
an extended rarefaction fan, a thin dense zone behind the CD
and an extended outer shell. Note the much higher bulk Lorentz factor near 
the leading edge of the outer shell, in the adiabatic case.

The light-curves generated during the fireball--EM collision were computed 
using a simplified energy release prescription which should give
the main features expected from a more realistic
radiation production treatment. We assumed that the lab-frame time-scale
for radiating the internal energy is $0.25\;t_{dec}$, in the sense defined 
above, and that radiation is emitted isotropically in the comoving 
frame. As shown by Rees (1996), the surfaces emitting photons that reach the 
observer simultaneously, are prolate ellipsoids with an axis ratio equal to
the bulk Lorentz factor of expansion of the radiating medium. Due to strong 
beaming effects, only those regions moving at an angle less than $\Gamma^{-1}$ 
relative to the axis origin--observer give a significant 
contribution to the light the detector receives.
This corresponds to half of the surface of the above 
mentioned ellipsoids and, because the ellipsoids are very elongated, the light 
received by the observer at any given moment comes from a surface which can be
extended over as much as one deceleration radius. Figure 4 (upper graph)
shows the light-curve obtained by integrating the radiation emitted by a burst 
1 Gpc away from the observer, without taking into account cosmological 
effects. The inner curves represent the flux generated at times less than
$t=0.5\;t_{dec}$ (the innermost curve), $t=0.6\;t_{dec}$, $t=0.7\;t_{dec}$ and 
so on up to $t=2.0\;t_{dec}$ (the outermost curve), in steps of $0.1\;t_{dec}$.
This allows one to see when photons emitted at a certain time during the
fireball--EM interaction arrive at the detector: light emitted before
$t=1.2\;t_{dec}$ contributes to the GRB peak, while the energy released after 
$t=1.2\;t_{dec}$ forms the tail. In this case the burst lasts 
$\approx 500$ seconds and has a peak at $T_p\simeq 80$ seconds. 
Most of GRB's light (92\,\%, see figure 4, middle graph)
comes from the shocked EM (the main heat store). The log-log 
light-curve in figure 6 (upper graph) shows that the burst rise is 
a 1.2 power-law in the range $40-80$ seconds, and that its tail (often 
referred to as ``exponential decay") can be well approximated by a $-1.2$ 
power-law in the range $100-1000$ seconds.

It is useful to compare these numerical results with the analytical calculations
of Fenimore et al. (1996). They obtained impulsive GRB light-curves assuming that:
(1) at any time, the radiating shell can be approximated as an infinitesimally thin 
surface, (2) the shell radiates at a constant comoving power $P_0$ from 
$t_0=2\,\Gamma^2\,T_0$ until $t_{max}=2\,\gamma^2\,T_{max}$ and is zero otherwise, and
(3) the shell moves with a Lorentz factor that is a power-law in the detector time $T$ :
$\gamma=\Gamma\,(T/T_0)^{-\zeta}$. Fenimore et al. (1996) derived that the photon fluxes 
(photons/${\rm cm}^2$s) should rise slower than $T^2$ and decay steeper than 
$T^{-2}$. The above assumptions are useful in order to develop an analytical model,
but they imply substantial departures from the situation represented by our
numerical results. In our case, the emitting 
structure is several hundreds light-seconds thick after $t=1.4\;t_{dec}$, 
so that the light emitted by the gas moving toward the observer
extends over the whole tail of the light-curve and, consequently, 
the structure cannot be approximated as infinitesimally thin. We do consider that 
radiation is emitted for a finite time (from $t_0=0.4\;t_{dec}$ to $t_{max}=
2.0\;t_{dec}$), but at a power that is not constant: it rises, has a 
peak around $t=1.0\,t_{dec}$ and then falls (figure 5, middle graph).
Moreover, different 
parts of the shocked structure move with different velocities. An averaged flow 
Lorentz factor, defined as the non-thermal energy--to--mass ratio, cannot
be approximated as a power-law in $T$ over the 
whole GRB history, but only locally.  Nevertheless, one can try to ignore these 
complications and use the assumptions listed above to compute analytically 
the GRB bolometric flux (ergs/${\rm cm}^2$s) and compare it with the numerical 
results. Similar to the derivations in Fenimore et al. (1996), we find that
the energy flux at detector is:
\begin{equation}
F(T)=\frac{8}{3\left(1-\zeta\right)}\frac{c^2}{D^2}\,P_0\,\Gamma^6\,T_0^{6\zeta}\,T^{-4} \times 
\left\{ \begin{array}{ll}
0 & T \leq T_0 \\
T^{6-6\zeta}-T_0^{6-6\zeta} &  T_0\leq T \leq T_{max} \\ 
T_{max}^{6-6\zeta}-T_0^{6-6\zeta} & T_{max} \leq T \\
\end{array} \right. .
\label{analyticlc}
\end{equation}  

One difficulty arises from the fact that the shell is not very thin
and $T_{max}$, defined as the time when light emitted by the fluid moving
directly toward the observer
reaches the detector, is not well defined: radiation
emitted at $t=t_{max}$ is spread in the light-curve over more than 500 seconds. 
If $T$ is the time when light emitted from the forward shock arrives at the 
detector, then $T_{max}=400$ s, while if the CD is used to 
define $T_{max}$, then $T_{max}=910$ s. Since the shell is very thin at $t_0=
0.4\;t_{dec}$, $T_0$ has an unique value of $T_0=17$ s. The power-law exponent 
$\zeta$ for the average Lorentz factor is between 0.15 and 0.28 if 
17 s $\leq T \leq$ 30 s, and between 0.53 and 0.61 if 100 s $\leq T \leq$ 1000 s,
depending on whether the blast wave or the CD is used to 
define $T$. Then, according to equation (\ref{analyticlc}), the light-curve 
should rise as $T^{\alpha}$ with $0.3\leq \alpha \leq 1.1$ for $T \simg$ 25 s, 
and should decay as $T^{-\alpha}$ with $1.2 \leq \alpha \leq 1.8$ for 
300 s $\leq T \leq$ 1000 s, and much steeper (as $T^{-4}$) for $T >$ 1000 s. 
Thus we obtained some analytical results that are close to the numerical ones
(see the upper graph in figure 6).
We emphasize the fact that the rise of the burst and the most important part
of its tail are in the $T_0 < T < T_{max}$ regime in equation (\ref{analyticlc}), 
and that only the much steeper decay at $T > 5000$ s in figure 6 corresponds 
to the  $T^{-4}$ fall for $T > T_{max}$ in equation (\ref{analyticlc}).

The kinetic, internal and radiated energies, normalized to the total initial 
energy $E=10^{51}$ ergs, and the radiated power in units of $E/t_{dec}=1.19 
\times 10^{45}\,\rm{ergs/s}$ are shown in figure 5 (upper and middle 
graphs). For the $0.25\;t_{dec}$ energy release time-scale considered, 87\,\% 
of the initial kinetic energy was converted into internal energy until $t=2.0\;
t_{dec}$, from which more than 96\,\% was released as radiation. As mentioned 
before, in this case the burst duration is determined by the how fast the  
kinetic energy is transformed into heat, i.e. the deceleration time-scale. Had 
we chosen a time-scale for energy release larger than the dynamical time-scale, 
the burst duration would have been determined by how fast the structure is able to 
radiate its internal energy, and the burst would last longer, having a 
lower fluence if a substantial fraction of the internal energy is lost 
adiabatically.

\section{Discussion}

For given parameters $E$, 
$r_0$, $\eta$ and $n$, the shape of the light-curves is basically  
independent of the details of the energy release mechanism, provided 
that the time-scale for radiating this energy is shorter than the deceleration 
time-scale (they should also be
similar if the energy release time-scale exceeds the dynamic time-scale, 
but the two scale similarly with radius; otherwise the light-curves would 
be affected by an additional r-dependent efficiency factor $t_{dyn}/t_{rad}$).
For a homogeneous EM, the burst is smooth, has a sharp rise and a
long power-law tail (rather than exponential). 
Figure 6 (top) is a log-log plot of the light-curves for 
$E=10^{51}$ ergs, $r_0=10^8$ cm, $n=1\;{\rm cm}^{-3}$ and different 
initial Lorentz factor $\Gamma$ (or parameter $\eta$): 50, 100 and 200. 
It can be noticed that, as $\Gamma$ is increased, the light-curve is shifted in 
the log F--log T plane, but its shape remains almost unchanged. Furthermore, 
the shift is the same for both pairs of successive parameters $\eta$. 
The peak fluxes scale as $\eta^{8/3}$ and the times $T_p$, $T_1$ and $T_2$ at
which the peak flux, and a fraction $e^{-1}$ or $e^{-2}$ of the peak flux are
reached, scale as $\eta^{-8/3}$ (this is in fact what is expected from eq.
[\ref{duration}]). A similar result is obtained if $\eta$ and $E$ are held 
constant and the EM density $n$ is changed or if $\eta$ and $n$ are fixed and 
$E$ is varied (see figure 6, bottom graph).
The peak fluxes scale as $E^{2/3}\,n^{1/3}$ while $T_p$, $T_1$ and 
$T_2$ go as $E^{1/3}\,n^{-1/3}$. We conclude that, with a good accuracy, the 
light-curves can be parameterized in $E$, $\eta$ and $n$ and reduced to a 
universal function $f(\tau)$:
\begin{equation}
F(E,\eta,n;T)=E^{2/3}\,\eta^{8/3}\,n^{1/3}\;f(\tau)\;, \quad
{\rm where} \quad \tau=E^{-1/3}\,\eta^{8/3}\,n^{1/3}\,T.
\end{equation}
This important feature of the external shock burst light-curves can be obtained 
analytically by substituting $T_0 \, (\propto t_0/\eta^2 \propto t_{dec}/\eta^2) 
\propto E^{1/3}\,\eta^{-8/3} \,n^{-1/3}$
(eq. [\ref{duration}]) in equation 
(\ref{analyticlc}), for $T_0 \leq T \leq T_{max}$ (which represents most of the 
light-curve rise and fall). Then the function
$f(\tau)$ is proportional to $P_0\,\eta^{-2}\,n^{-1}\;\tau^{2-6\zeta}$, 
which implies that $P_0$ must be proportional to $\eta^2\,n$, in order to remove 
the multiplying factor dependence on burst parameters. This is indeed the case, 
since we assumed that the comoving radiated power is proportional to the 
comoving internal energy density of the shocked gas, which \cite{blamc76} show 
to be proportional to $\eta^2\,n$. Nevertheless, it must be emphasized that the 
assumptions used in the analytic derivations are rough, and the most 
reliable proof for the existence of such a simple scaling relation for $F(T)$ 
is based on the numerical results.

The effects of an inhomogeneous EM on the dynamics and energetics of the collision
and on the light-curve can
be estimated using a simple model of an expanding fireball interacting
with an EM whose density varies as power-law (also of interest for
the case where the event occurs inside a pre-ejected wind or nebula). Figure
4 (lower graph) shows the light-curves generated by the collision with
an EM whose density is increasing as $r^{2.5}$ or decreasing as $r^{-1}$,
starting from $r=0.4\,r_{dec}$ (where $n=1 \,{\rm cm}^{-3}$),
compared with the homogeneous EM case. It can be seen that the light-curve is
very insensitive to a monotonously changing EM density. The reason for this can
be understood from figure 5 (middle and lower graphs), representing the
shocked structure radiating power and average Lorentz factor as functions of
the radial coordinate (the deceleration time-scale $t_{dec}$ and radius $r_{dec}$
are those previously defined for a homogeneous EM). When the external density
increases with radius, the shocked gas on either side of the CD
radiates more efficiently than in the homogeneous case, but the
average Lorentz factor is lower and the emitted light is more
stretched in detector time T than in the homogeneous EM case. The two effects act
against each other, resulting in a light-curve that peaks at almost the same
time and flux as for a homogeneous EM. A similar situation occurs when the
external density decreases with $r^{-1}$ (or with $r^{-2}$, as in a constant
velocity wind): the two shocked regions radiate less efficiently, but the average Lorentz
factor is larger, generating in the detector a flux with a very similar
T-dependence. The impact of a fireball on an inhomogeneous, blobby medium
(\cite{fenetal96}) would therefore lead to similarly shaped light-curves,
with individual sub-pulses having the same generic shape as the light-curves in
a homogeneous medium. We conclude that bursts arising from external shocks have
a fast rise and a long decay, and are insensitive to variations in the EM density,
as long as the energy release time-scale is shorter than the deceleration time-scale.

\cite{link96} studied the
temporal asymmetry of over 600 bursts from the BATSE 3B catalog, 
using the third moment of the burst time profile, and found
that about two-thirds of them have positive time-asymmetry, in the sense that the flux
rises faster than it falls. \cite{nemiroff} used the brightest GRBs 
longer than 1 second detected by BATSE up to March 1993,
and found that there are significantly more bursts that rise faster
than the decay, on all energy bands. 
The average of the ratio fall--to--rise times over the full sample
of bursts was found to increase with the time bin size, from
1.1 on the 64 ms time-scale to 1.4 on the 4096 ms time-scale.
Mitrofanov et al. (1996) used 338 events of the Second BATSE catalog, that are longer
than 1 second, aligned the normalized fluxes of all bursts at the peak time,
and calculated the ``averaged curve of emissivity" (ACE). Such an averaged
light-curve represents the general time history of GRBs and is appropriate for 
a comparison with our numerical results (single hump light-curves), since its
construction averages out
all individual peaks within bursts and allows the calculation of a time-asymmetry
ratio that is not affected by the asymmetry of these peaks. Mitrofanov et
al. (1996) computed the rise time $T_R$ and fall time $T_F$ of the ACE by 
adding the flux weighted time bin durations 
in the ACE's rise and fall, respectively. 
They found that $T_R=2.33\,{\rm s}$ and $T_F=4.21\,{\rm s}$, which gives an asymmetry
ratio $T_F/T_R=1.81$. We note that the fall of the ACE (shown in figure 2 in their
article) can be fit by a power-law of index -1.0, with time measured from $T_p-1.87\,{\rm s}$,
where $T_p$ is the peak time.

Using equation (\ref{analyticlc}) and the light-curve for $\Gamma=100$
(figure 4, top graph), we find that the fall of the ACE published by Mitrofanov et al.
(1996) can be best fit by the tail of the light-curve generated from an external shock with
$\Gamma\approx 400$. The time-scales and asymmetry ratio of this burst are
$T_R=0.90\,\rm{s}$, $T_F=4.03\,\rm{s}$ and $T_F/T_R=4.50$. Its fall is a -1.16 power-law
in time measured from $T_p-2.06\,{\rm s}$. In conclusion, our model can reproduce well
the tail of the ACE mentioned above, but the rise of the numerical light-curve is
faster by a factor $\sim 2.6$, leading to a burst that is
more asymmetric than the reported observations. Moreover, the external shock model
with constant energy release time-scale always gives profiles with a decay longer than the rise,
whereas a small
fraction of the observed bursts shows the opposite trend. This implies either that 
additional factors are contributing to the light-curves from external bursts 
(e.g. perhaps a radiative precursor propagating ahead of the shock), or a 
significant fraction of bursts arise from internal shocks where the light-curve 
is dictated by the details of an extended energy release that generated the fireball
(\cite{rm94}; \cite{fenetal96}; \cite{pm96}). The latter conclusion is also reinforced by the 
fact that internal shock bursts would be in principle able to show not only 
very different light-curve envelopes but also short time variability, whereas 
the external shock light-curves calculated here result in smooth profiles.
This is due essentially to the smearing introduced by the
simultaneous reception of radiation from regions moving at Lorentz factors $\gamma$ 
that differ by factors $\sim 2$. 
Nevertheless, our model should generate 
less asymmetric light-curves, and even ones with a rise front lasting
longer than the tail, if specific energy release processes are taken 
into account and if these processes have time-scales 
that are longer than the dynamical time-scale in the
beginning of the shell--EM collision ($t < t_{dec}$), and decrease as
the interaction progresses, becoming shorter than $t_{dec}$ at later
times ($t > t_{dec}$).

In conclusion, we have developed a one-dimensional, finite difference + 
Riemann solver hydrodynamic code which is able to simulate the propagation
of shocks with Lorentz factors of few hundreds and to evolve 
the shocked structure over a time that is $\sim 10^4\,\Gamma_2^2$ times longer 
than its lab frame crossing time.
We have used this code to simulate external shocks in cosmological Gamma-Ray Burst 
models arising from an initial impulsive energy release, and to analyze 
the dynamics of the forward and reverse shocks, as well as the shock
and rarefaction wave reflections.
These computations confirm the general conclusions from previous simplified 
analytic descriptions of shock dynamics and light-curves, and 
provide detailed quantitative information 
which can be employed for comparisons with the observational data.
We have also found a simple scaling law for bursts generated by external
shocks with different physical parameters.

\acknowledgements{This research has been supported in part through NASA 
NAG5-2362, NAG5-2857, NSF-PHY 93-09834 and NSF-PHY 93-57219}

\pagebreak

\input psfig

\begin{figure}
\centerline{\psfig{figure=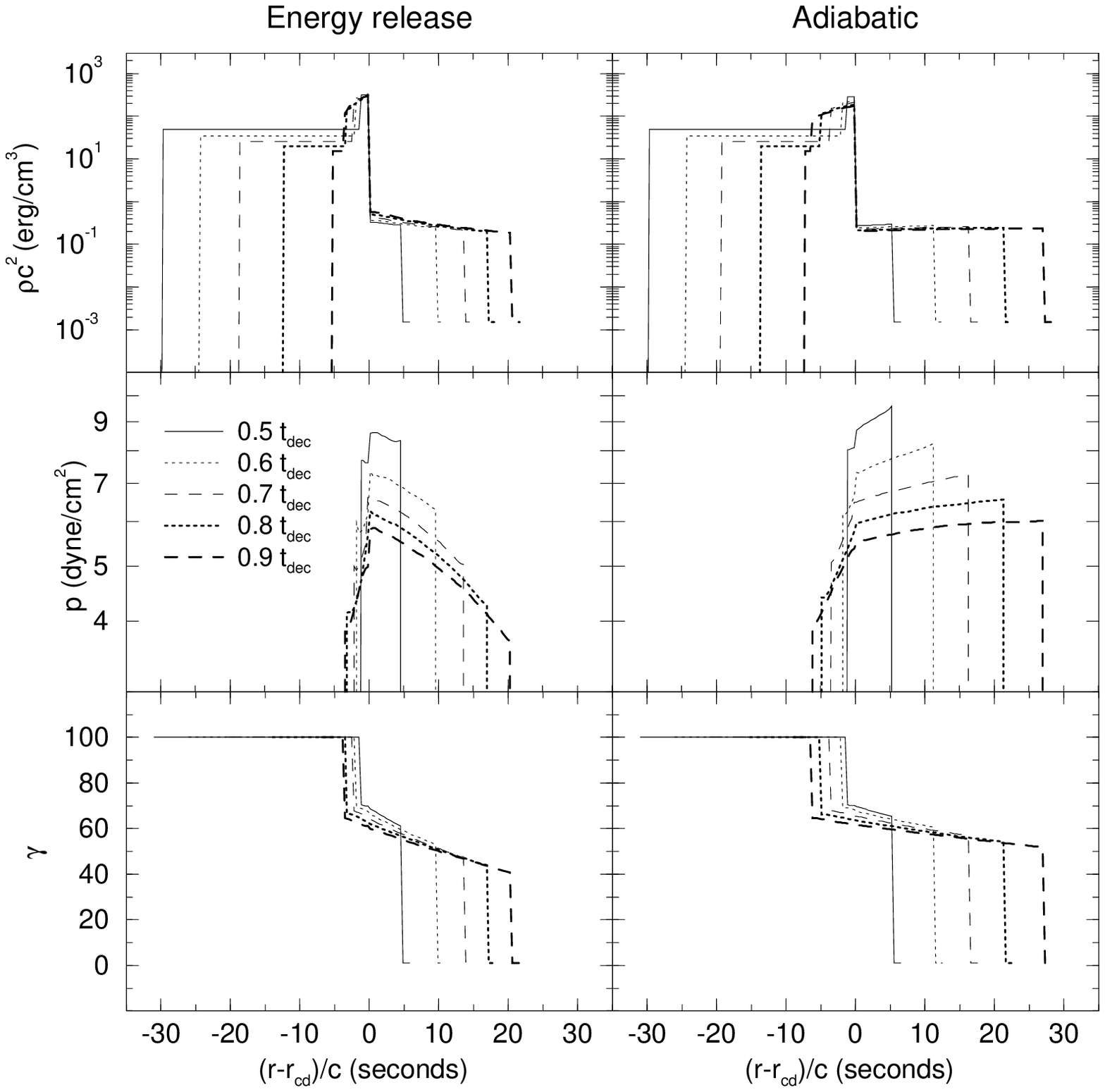}}
\vspace{1 in}
\figcaption{Density, pressure and flow Lorentz factor for $\Gamma=100$, $E=10^{51}$ ergs, 
$r_0=10^8$ cm and $n=1\;{\rm cm}^{-3}$, at times indicated in the legend.
The left column shows these profiles for the $0.25\,t_{dec}$ energy release time-scale,
while the right column is for the adiabatic interaction. The structure is much thinner than its
curvature radius and the position inside it is indicated relative to the contact discontinuity.
Negative values correspond to the inner shell, positive values to the outer shell. 
Note that in the adiabatic interaction, the outer shell is less dense and more extended, 
and that the gradients in density, pressure and Lorentz factor are smaller.
After $t=0.9\;t_{dec}$ the reverse shock crosses the inner shell, in both cases.}
\end{figure} 

\begin{figure}
\centerline{\psfig{figure=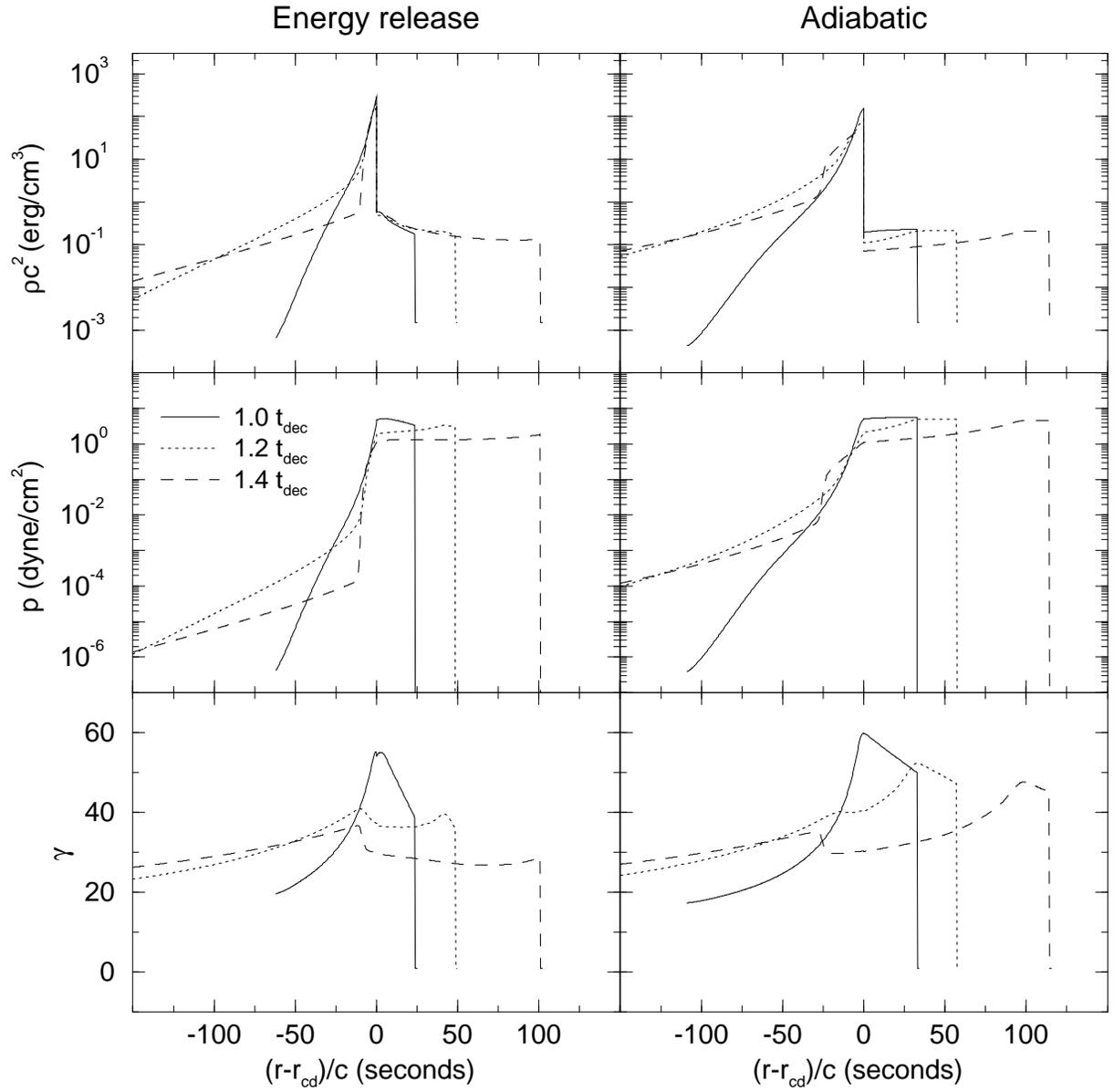}}
\vspace{1 in}
\figcaption{The same profiles in the energy release and adiabatic cases,
 after the reverse shock has crossed the fireball. 
The second reverse shock can be seen easier in the  Lorentz factor $\gamma$ graph.} 
\end{figure}

\begin{figure}
\centerline{\psfig{figure=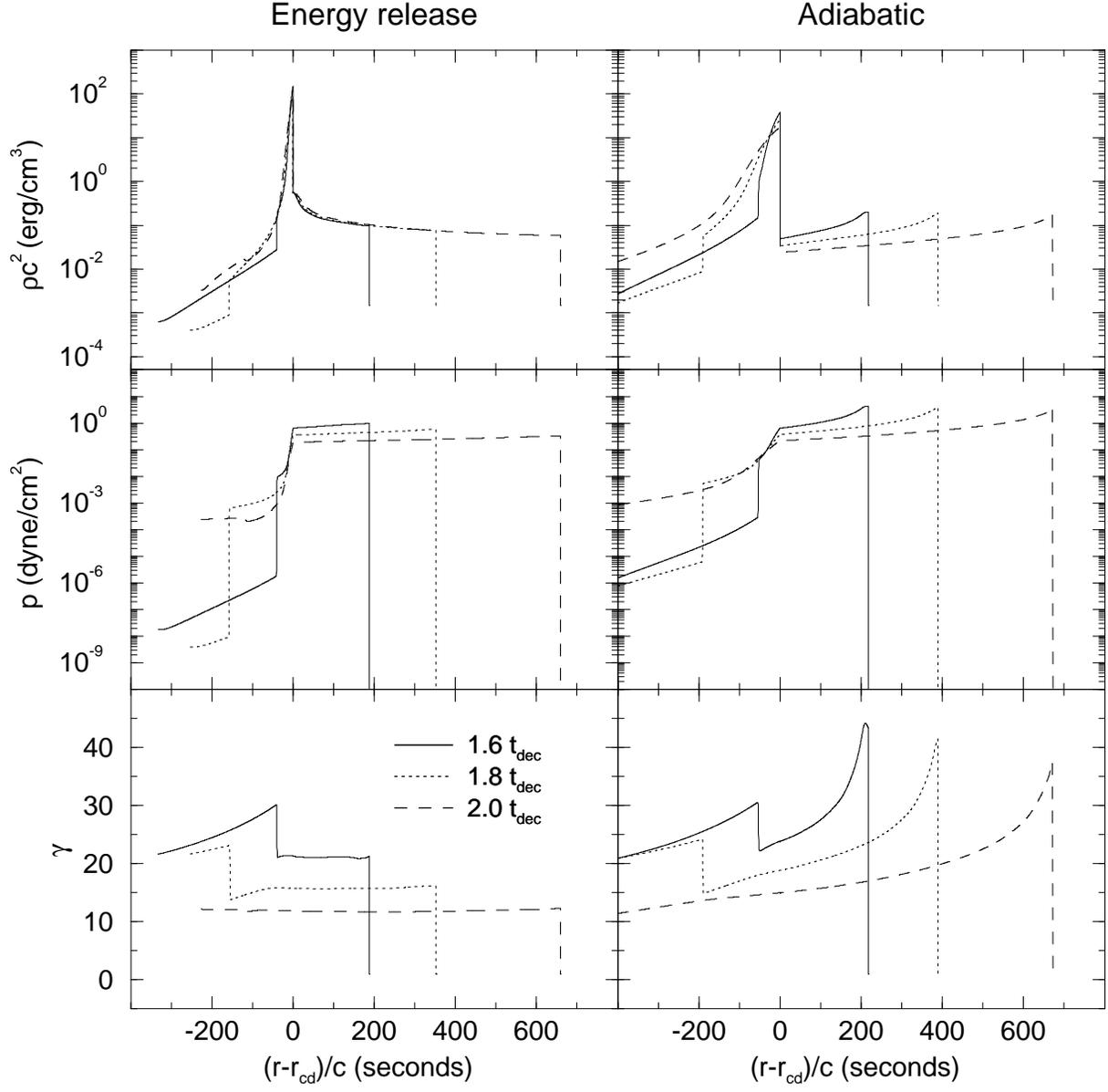}}
\vspace{1 in}
\figcaption{Same as in figures 1 and 2, until $t=2.0\;t_{dec}$,
when most of the initial kinetic energy has been radiated (in the non-adiabatic case) 
and the simulation is ended. Shortly before $t=2.0\;t_{dec}$, the second reverse shock crosses
the rarefaction fan behind the contact discontinuity.}
\end{figure}

\begin{figure}
\centerline{\psfig{figure=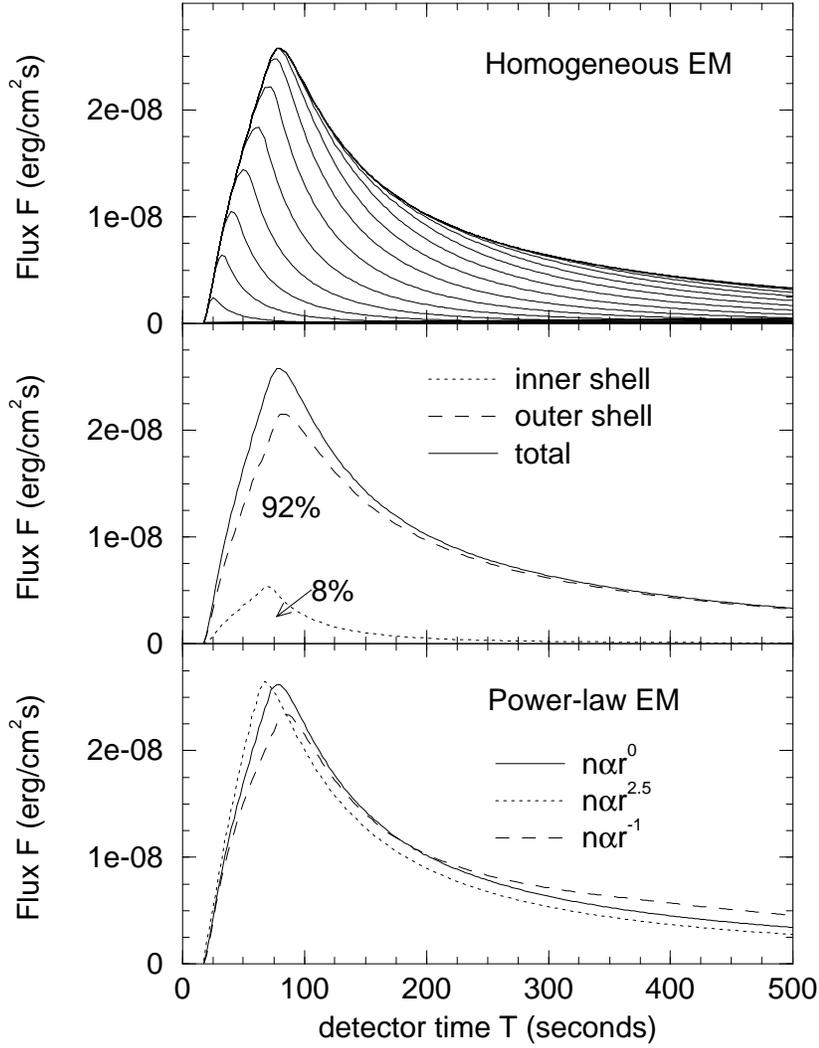}}
\vspace{1 in}
\figcaption{Upper graph: light-curve for $\Gamma=100$ and $0.25\;t_{dec}$ energy release
time-scale. The innermost curve is the light emitted by the shocked matter until 
$t=0.5\;t_{dec}$, while the envelope is for the light emitted until $t=2.0\;t_{dec}$. 
Between these two curves are shown the light-curves generated until $t=0.6\;t_{dec}$,
$t=0.7\;t_{dec}$, $t=0.8\;t_{dec}$ ..., in $0.1\;t_{dec}$ steps.
The light-curve peaks at $T_p\simeq 80$ seconds. Middle graph: the shocked fireball
and shocked EM contributions to the total light-curve. Note the large difference 
between the energies radiated by the two shocked media. Lower graph: comparison
between the light-curves generated by a homogeneous EM (solid curve) and a power-law
EM (dotted and dashed curves).} 
\end{figure}

\begin{figure}
\centerline{\psfig{figure=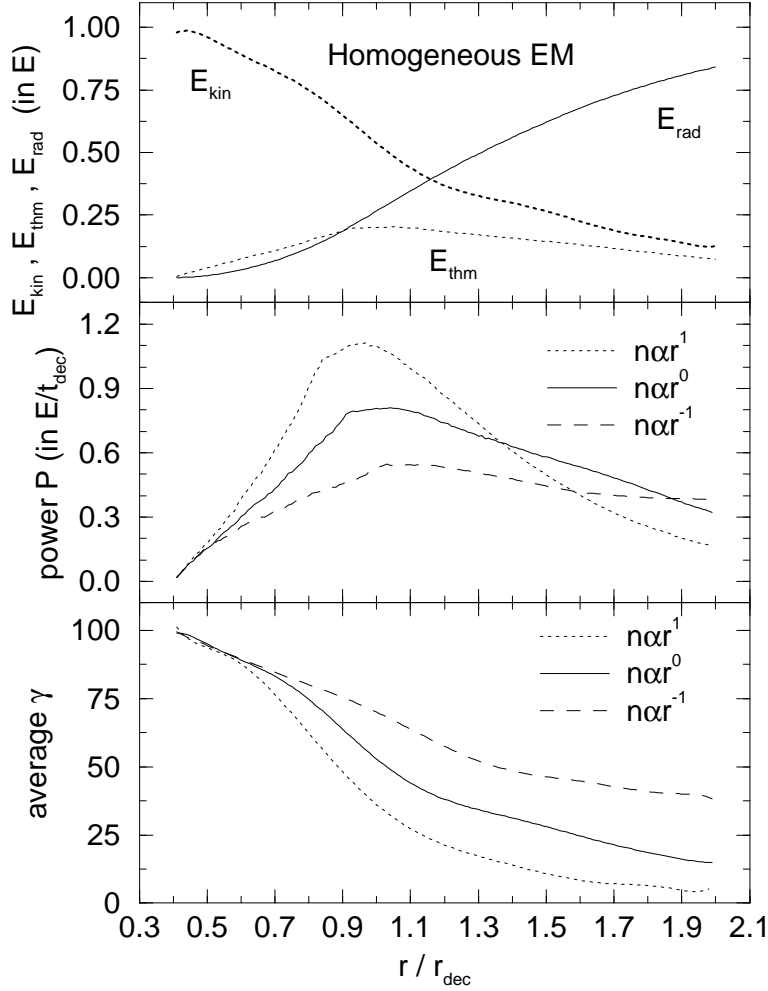}}
\vspace{1 in}
\figcaption{Upper graph: Time evolution of kinetic (thick dotted line), internal 
(thin dotted line) and radiated energy (thin solid line), as the structure travels from
$0.4\;r_{dec}$ to $2.0\;r_{dec}$. All energies are normalized to the initial kinetic energy
of the fireball. Middle graph: radiating power in units of $E/t_{dec}$ for homogeneous
and power-law EM density. $t_{dec}$ and $r_{dec}$ as defined for a homogeneous EM are
used, to allow comparison. The radiated power peaks earlier if the EM has an increasing
density, as one would expect. Lower graph: average Lorentz factor for the same EM densities. 
$E=10^{51}$ ergs, $r_0=10^8$ cm and $\eta=100$.} 
\end{figure}

\begin{figure}
\centerline{\psfig{figure=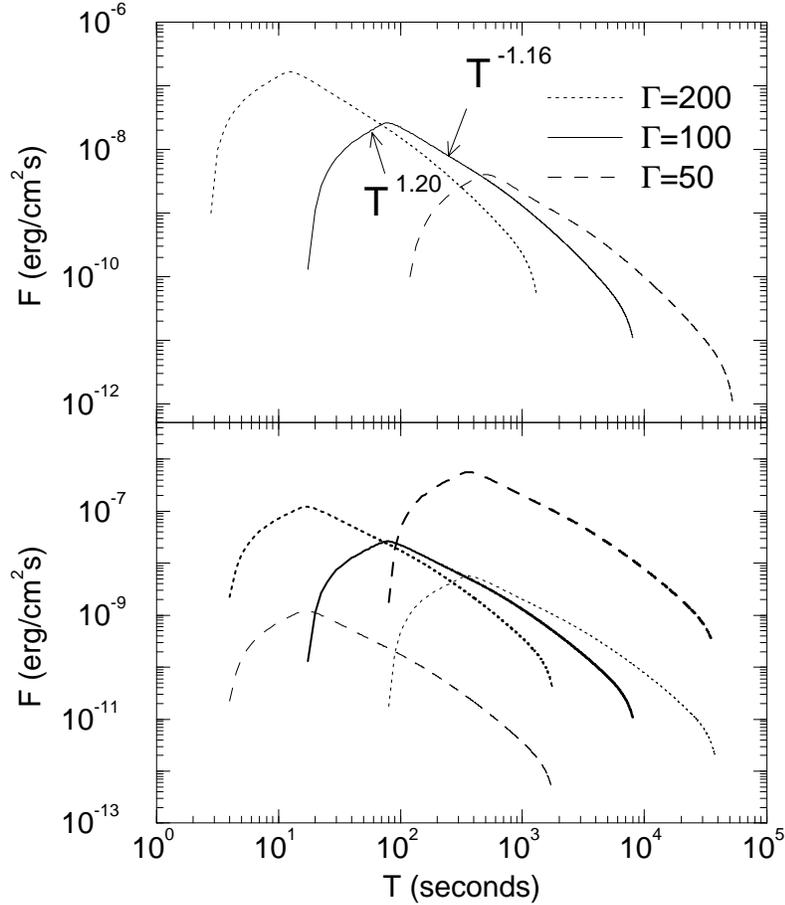}}
\vspace{1 in}
\figcaption{Upper graph: light-curves for $E=10^{51}$ ergs, $r_0=10^8$ cm, $n=1\;{\rm cm}^{-3}$ and
different initial Lorentz factors $\Gamma$ (or parameters $\eta$). 
Note that all light-curves have the same shape and
shift equally when $\Gamma$ is doubled. Two regions, were the received flux is a power-law in
the detector time $T$, are identified. Lower graph:
light-curves for $\eta=100$, $r_0=10^8$ cm and : $E=10^{51}$ ergs, $n=1\,{\rm cm}^{-3}$
(thick solid curve); $E=10^{51}$ ergs, $n=0.01\,{\rm cm}^{-3}$ (thin dotted curve);
$E=10^{51}$ ergs, $n=100\,{\rm cm}^{-3}$ (thick dotted curve); $E=10^{49}$ ergs, $n=1\,{\rm cm}^{-3}$
(thin dashed curve); $E=10^{53}$ ergs, $n=1\,{\rm cm}^{-3}$ (thick dashed curve).}
\end{figure}

\end{document}